\documentclass[preprintnumbers, pra, showpacs, floatfix,twocolumn,
preprintnumbers, letterpaper, superscriptaddress]{revtex4}
\usepackage{amsfonts}
\usepackage{amsmath}
\usepackage{amssymb,epsf}
\usepackage{latexsym}
\usepackage{graphicx,epsfig}
\usepackage{amssymb}
\usepackage{subfigure}
\usepackage[colorlinks=true,citecolor=blue,linkcolor=blue,urlcolor=black]{hyperref}
\usepackage{epstopdf}
\usepackage{color}

\def\be{\begin{equation}}
\def\ee{\end{equation}}
\def\ba{\begin{eqnarray}}
\def\ea{\end{eqnarray}}
\def\la{\langle}
\def\ra{\rangle}

\begin{document}
\title{Kosterlitz-Thouless phase and $Z_d$ topological quantum phase }
\author{Mohammad Hossein Zarei}
\email{mzarei92@shirazu.ac.ir}
\affiliation{Physics Department, College of Sciences, Shiraz University, Shiraz 71454, Iran}

\begin{abstract}
It has been known that encoding Boltzmann weights of a classical spin model in amplitudes of a many-body wave function can provide quantum models whose phase structure is characterized by using classical phase transitions. In particular, such correspondence can lead to find new quantum phases corresponding to well-known classical phases. Here, we investigate this problem for Kosterlitz-Thouless (KT) phase in the d-state clock model where we find a corresponding quantum model constructed by applying a local invertible transformation on a d-level version of Kitaev's Toric code. In particular, we show the ground state fidelity in such quantum model is mapped to the heat capacity of the clock model. Accordingly, we identify an extended topological phase transition in our model in a sense that, for $d \geq 5$, a KT-like quantum phase emerges between a $Z_d$ topological phase and a trivial phase. Then, using a mapping to the correlation function in the clock model, we introduce a non-local (string) observable for the quantum model which exponentially decays in terms of distance between two endpoints of the corresponding string in the $Z_d$ topological phase while it shows a power law behavior in the KT-like phase. Finally, using well-known transition temperatures for d-state clock model we give evidences to show that while stability of both $Z_d$ topological phase and the KT-like phase increases by increasing $d$, the KT-like phase is even more stable than the $Z_d$ topological phase for large $d$.

\end{abstract}
 \maketitle
\section{Introduction}
Characterizing different phases of matter is a central problem in condensed matter physics \cite{sav}. While this problem seems well established in classical physics, it is specifically challenging in quantum physics where a completely different property namely entanglement plays a very specific role \cite{entan}. This problem has led to a cross-fertilization between condensed matter physics and quantum information theory where using concepts from quantum information theory, one will be able to characterize a quantum phase transition \cite{hamma, chen, mont2010}. In particular, besides different measures provided by quantum information theory, the ground state fidelity has attracted much attention during past decades where a quantum phase transition can well be characterized by a singularity in the ground state fidelity \cite{fd,gu,fidelity}.

Among different quantum phases, topological quantum phases have attracted much attention because of their applications in fault tolerant quantum computation. A well-known example is Kitaev's toric code \cite{Kitaev2003} which shows a robust topological degeneracy against any local perturbation \cite{rob1, rob2, zare16}. The abelian d-level version of this model, which we call a $Z_d$ Kitaev model, is also interesting where it has been known as a string-net model with a $Z_d$ topological phase \cite{stringnet,zd,vidal1,vidal2} which is even more robust than of $Z_2$ one \cite{del,karimi}.  In spite of the above important applications, since topological phases have a non-local order and do not follow the symmetry breaking paradigm of Landau \cite{land}, their characterization is a challenging task. Different approaches for characterizing topological phases\cite{wwen,wang,wen3,kitpre} are based on a fact that non-local nature of a topological order should lead to a stability against any local transformation. In particular, considering local stochastic transformations (i. e. local invertible (LI) transformations) on entangled states has recently been known as a very important approach where it has been shown that topological phases are stable against small  LI transformations \cite{wen4}.

On the other hand, recently it has been common to consider some quantum phase transitions by mapping to classical phase transitions \cite{castel2005,r1,r3,Somma2007,zarei18,Dennis2002,Katzgraber2009,zareim19,zareiab,fer,fer1}. An interesting idea behind such mappings is that the thermal fluctuations are mapped to quantum fluctuations by encoding Boltzmann weights of a classical model in amplitudes of a quantum entangled state. In particular, it has been shown that the ground state fidelity in such entangled states is mapped to the heat capacity of the corresponding classical models \cite{castel, zarei19}. Therefore, any singularity in the heat capacity corresponds to a singularity in the ground state fidelity and the corresponding quantum and classical phase transitions will be related to each other.

Motivated by the above classical-quantum mappings, one can ask is it possible to find new quantum phases by considering such mappings for various classical spin models with well-known classical phases? In particular, one of the most interesting classical phases is Kosterlitz-Thouless (KT) phase which has originally been seen in classical X-Y model \cite{kt}. Such a phase has also been seen in classical d-state clock model where, for $d \geq 5$, a KT phase emerges between ferromagnetic and paramagnetic phases \cite{clock00, clock1,clock0,clock2}. In this respect, in order to find a quantum phase corresponding to the KT phase, it is enough to find a quantum entangled state correponding to the classical clock model. Interestingly, we show that such quantum state is a simple deformation of $Z_d$ Kitaev state where it is constructed by applying an LI transformation in the $Z_d$ kitaev state and can also be considered as the ground state of a quasi-hermitian Hamiltonian \cite{bender1,bender2,mostafa,quasi1}. We explicitly prove that the ground state fidelity of such quantum model is mapped to the heat capacity of the clock model. Therefore, we identify an extended topological phase transition in the deformed Kitaev model where a KT-like quantum phase emerges between the $Z_d$ topological phase and a trivial phase.

We also give a mapping from the correlation function in the clock model to a non-local (string) observable \cite{string} in the quantum model. Accordingly, we show that such string parameter can characterize the KT-like phase in a sense that it exponentially decays in terms of distance between two endpoints of the corresponding string in the $Z_d$ topological phase while it shows a power law behavior in the KT-like quantum phase. Furthermore, by the fact that  for $ d \leq 4$ clock model shows a single ferromagnetic-paramagnetic phase transition, we conclude that in this case the KT-like quantum phase disappears in the quantum model. Finally, we also use the well-known transition temperatures for the clock model for different values of $d$ \cite{clock00,clock1,clock0} to derive transition points of our quantum model. In particular, by an interpretation of transition points as measure of stability against small LI transformation, we give evidences which show that stability of both $Z_d$ topological phase and KT-like phase increases by increasing $d$. Furthermore, we show that the KT-like phase is even more stable than the $Z_d$ topological phase for large $d$.

This paper is structured as follows: In Sec.(\ref{sec1}), we give a brief review on the $Z_d$ Kitaev state and it's corresponding Hamiltonian. In Sec.(\ref{sec2}), We introduce the deformed Kitaev model constructed by an LI transformation on the $Z_d$ Kitaev model. Then in Sec.(\ref{sec3}), we prove the correspondence between the ground state fidelity and the heat capacity and finally in Sec.(\ref{sec4}), we give main results of the paper where we use well-known facts about the clock model to identify the KT-like quantum phase.

 \section{$Z_d$ Kitaev state}\label{sec1}
\begin{figure}[t]
\centering
\includegraphics[width=7cm,height=5cm,angle=0]{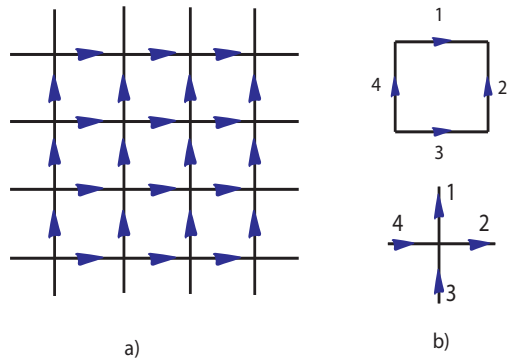}
\caption{a) A square lattice with qudits living in edges. each edge is labeled by a direction. b) A plaquette operator of $Z_1 Z_2 ^{-1} Z_3 ^{-1} Z_4$ and a vertex operator of $X_1 X_2 X_3 ^{-1} X_4 ^{-1}$ are defined corresponding to each plaquette and vertex of the lattice, respectively. } \label{d-kitaev}
\end{figure}

$Z_d$ Kitaev state is defined as a simple generalization of Kitaev's Toric code state where qubits are replaced by d-level quantum systems called qudits \cite{qudit1,qudit4,qudit2,qudit3}. Therefore, consider a two dimensional $L \times L$ square lattice with qudits living in edges of the lattice. Furthermore, we also label each edge by a direction, for example see Fig.(\ref{d-kitaev}-a).  Like qubit case, $Z_d$ Kitaev state is also a stabilizer state which is stabilized by a generalized stabilizer group on $N$ qudits. A generalized stabilizer group is a subgroup of generalized Pauli group $P_N$ which are constructed by product of d-level Pauli operators and commute with each other. D-level Pauli operator of $Z$ is a diagonal matrix with eigenvalues in the form of $1, \omega, \omega ^2 , ...\omega ^{(d-1)}$ where $\omega =\exp \{2\pi i/d \}$ in a sense that if we denote eigenstates of the $Z$ by $|m\ra$, we will have $Z=\sum_{m=0}^{d-1} \omega^m |m\ra \la m|$. The Pauli operator of $X$ is also a ladder operator in the form of  $X=\sum_m |m+1\ra \la m|$ and $X|m\ra =|m+1\ra$.  It is clear that these generalized Pauli operators are not hermitian but they are unitary in a sense that $XX^{-1}=1$ and $Z Z^{-1}=1$. Furthermore, since $\omega ^d =1$, one can conclude that $X^d =Z^d =1$. Finally, one can check that there is a commutation relation between $X$ and $Z$ operators in the form of $ ZX = \omega XZ$.

Now we are ready to introduce stabilizers of the $Z_d$ Kitaev state. To this end, consider a plaquette of the lattice as shown in Fig.(\ref{d-kitaev}-b). Corresponding to each plaquette of the lattice, a plaquette operator is defined in the following form:
\begin{equation}
B_p =Z_1 Z_2 ^{-1} Z_3 ^{-1} Z_4
\end{equation}
If we turn around the plaquette in a clockwise direction, the above stabilizer can be  written in a general form as $\prod_{e \in \partial p} Z_e ^{\sigma_e}$ where $e \in \partial p$ refers to edges living around the plaquette $p$ and $\sigma_e$ is equal to 1 if direction of that edge is matched with the clockwise direction and is equal to -1 otherwise. The advantage of such general definition is that it is independent of directions that we had considered for edges in Fig.(\ref{d-kitaev}-a) where it works for any given direction for edges. Then, corresponding to each vertex of the lattice, see Fig.(\ref{d-kitaev}-b), a vertex operator is defined in the following form:
\begin{equation}
A_v =X_1 X_2  X_3 ^{-1} X_4 ^{-1}
\end{equation}
This operator can also be written in a compact form as $\prod_{e \in v} X_e ^{\gamma_e}$ where $e\in v$ refers to edges connecting to the vertex $v$ and $\gamma_e$ is equal to $-1$ if direction of the edge $e$ is incoming to the vertex $v$ and $+1$ is equal to $+1$ otherwise. According to commutation relation of generalized Pauli operators and since each vertex operator has a zero or two common qudits with plaquette operators, it is simple to check that vertex and plaquette operators commute with each other. Therefore, vertex and plaquette operators generate a stabilizer group.

Then, the $Z_d$ Kitaev state, denoted by $|K_d\ra$, is defined as a quantum state which is stabilized by vertex and plaquette operators where $B_p |K_d \ra =|K_d\ra$ and $A_v |K_d\ra =|K_d\ra$. Up to a normalization factor, such a state can be written in the following form:
\begin{equation}\label{eq1}
|K_d \ra =\prod_v (1+A_v +A_v ^2 +...+A_v ^{d-1})|0\ra^{\otimes N}
\end{equation}
where $N=2L^2$ is total number of qudits. In order to show that the above state is stabilized by $A_v$'s and $B_p$'s, it is enough to note that $A_v ^d =1$ and therefore we have $A_v (1+A_v +A_v ^2 +...+A_v ^{d-1})=(1+A_v +A_v ^2 +...+A_v ^{d-1})$. Furthermore, since $B_p$ commutes by $A_v$'s and $B_p |0\ra ^{\otimes N}=|0\ra ^{\otimes N}$, it is simply concluded that $B_p |K_d\ra=|K_d\ra$. Furthermore, in the same way, one can prove that the $Z_d$ Kitaev state can also be written in the following form, up to a normalization factor:
\begin{equation}\label{eq01}
|K_d \ra =\prod_p (1+B_p +B_p ^2 +...+B_p ^{d-1})|+\ra^{\otimes N}
\end{equation}
 where $|+\ra= \frac{1}{\sqrt{d}}(|0\ra +|1\ra +...+|d-1\ra)$ is eigenstate of Pauli operator $X$.

On the other hand, the $Z_d$ Kitaev state can also be considered as a ground state of a hermitian Hamiltonian in the following form:
\begin{equation}
H_0 =-\sum_p (B_P +B_P ^{-1}) -\sum_v (A_v +A_v ^{-1})
\end{equation}
Interestingly, it has been shown that when the square lattice has a periodic boundary condition on a torus, the above Hamiltonian shows a topological degeneracy where $|K_d \ra$ is only one of the ground states of the system. It is shown that other ground states can be constructed by applying a few non-local $X$-type operators corresponding to non-contractible loops around torus \cite{Kitaev2003}. Therefore, it is concluded that the above degeneracy is robust against any local perturbation. Such a robustness is in fact a universal signature of all topological phases.

\begin{figure}[t]
\centering
\includegraphics[width=5cm,height=5cm,angle=0]{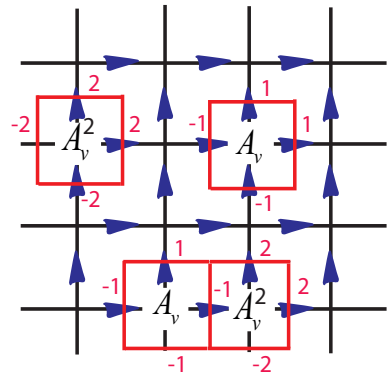}
\caption{A vertex operator is represented by a loop on the dual lattice where there is also a weight for each edge of the loop corresponding to power of $X$ operator in the $A_v$. $A_v ^2$ is also represented by a loop with different weights corresponding to power of $X$ in $A_v ^{2}$. A product of $A_v$ and $A_v ^{2}$ corresponding to two neighboring vertices is represented by a string-net where the weight of common edge of two initial loops is derived by a summation of the initial weights.} \label{p-v}
\end{figure}

From a microscopic perspective, it has been shown that topological order in $Z_d$ Kitaev state is related to a condensation of string-nets with a long-range entanglement. It is specially simple to see string-nets in Eq.(\ref{eq1}). To this end, note that if we expand the operator of $\prod_v (1+A_v +A_v ^2 +...+A_v ^{d-1}) $ where it is product on all vertices of the lattice, it will be a summation of all $X$-type stabilizers of the $Z_d$ Kitaev state which are constructed by product of $A_v$'s. On the other hand, each vertex operator of $A_v$ can be represented by a loop on the dual lattice as shown in Fig.(\ref{p-v}). However, since each power of $A_v$ in the form of $A_v ^m$ for $m=\{1,2,...,d-1\}$ is also a stabilizer, we should represent such stabilizers by weighted loops where the weight of each loop will correspond to power of $m$. Furthermore, a product of weighted vertex operators for different vertices will be also a stabilizer. Such stabilizers are also represented by product of weighted loops. In particular, if two weighted loops have a common edge, their product should be represented by a different structure in a sense that weight of the common edge will be equal to summation of weights of two initial loops, see Fig.(\ref{p-v}). Generally, such structures are called string networks and each stabilizer of the $Z_d$ Kitaev state corresponds to one of such string-networks. In this way, it is simple to see that the $Z_d$ Kitaev state is a superposition of string-networks and therefore it is called a string-net condensed state.
 \section{Local invertible transformation on the $Z_d$ Kitaev state: a deformed Kitaev state}\label{sec2}
 In this section, we introduce a deformation of $Z_d$ Kitaev state by applying an LI transformation, see also \cite{vidal1, vidal2} for similar deformations. To this end, consider an invertible transformation on a single qudit in the form of $\exp \{\frac{\beta}{2} (Z+Z^{-1})\}$ where $\beta$ is a positive real number. Then we consider an LI transformation as product of $\exp \{\frac{\beta}{2} (Z+Z^{-1})\}$ on all qudits of the $Z_d$ Kitaev state in the following form:
 \begin{equation}
 |K_d\ra~~~\rightarrow ~~~ \exp \{\frac{\beta}{2} \sum_i (Z_i +Z_i ^{-1})\}|K_d \ra
 \end{equation}
Since the invertible transformation is not unitary, it does not preserve the norm of the quantum state. However, we can add a normalization factor to the final state to have a normalized state denoted by $|K_d (\beta)\ra$. Note that for an arbitrary value of $\beta$ the above transformation can be considered as a sequence of small transformations which gradually map the initial state to the final state. In particular, it is interesting to consider the final state when $\beta \rightarrow \infty$. To this end, note that since $Z=\sum_{m=0}^{d-1} \omega^m |m\ra \la m|$, one can show that $\exp \{\frac{\beta}{2} (Z+Z^{-1})\}=\sum_{m=0}^{d-1} \exp \{\beta cos \frac{2\pi m}{d} \} |m\ra \la m|$ which can be written in the form of $\exp \{\beta \}(|0\ra \la 0| + \exp \{\beta(cos \frac{2\pi}{d} -1)\}|1\ra \la 1| +...+\exp \{\beta(cos \frac{2\pi (d-1)}{d}  -1)\}|d-1\ra \la d-1|$. Then, since $\cos x\leq 1$, for $\beta \rightarrow \infty$ the above operator converts to a projective operator of $|0\ra \la 0|$. By this fact, we conclude that the invertible transformation on the $Z_d$ Kitaev state converts it to a trivial state of $|000...0\ra$ for the limit of $\beta \rightarrow \infty$.

 In this way, it seems that, by an infinite sequence of small invertible transformations, we will be able to convert a topological state to a trivial state. It means that in space of quantum states on $N$ qudits, we have transition between two different quantum phases. On the other hand, it is clear that $|K(\beta)\ra$ can be considered as the ground state of a Hamiltonian which is constructed by applying the same LI transformation to the $H_0$ in the following form:
 \begin{equation}
 H_{\beta} = \exp \{\frac{\beta}{2} \sum_i (Z_i +Z_i ^{-1})\} H_0 \exp \{-\frac{\beta}{2} \sum_i (Z_i +Z_i ^{-1})\}
 \end{equation}
 Since the invertible transformation is a similarity transformation, it preserves the real energy spectrum of the initial $Z_d$ Kitaev model. Therefore, if we denote eigenstates of $H_0$ by $|K_i \ra$, eigenstates of $H_{\beta}$ will be in the form of $\exp \{\frac{\beta}{2} \sum_i (Z_i +Z_i ^{-1})\}|K_i \ra$ with the same eigenvalues. In particular, $|K_d (\beta)\ra$ is the ground state of the above Hamiltonian.

We should also emphasize that the $H_{\beta}$ is in fact a non-hermitian Hamiltonian with real eigenvalues and with eigenstates which are not orthogonal. However, it is not an important problem because it has been shown that one can define a different metric for definition of inner product in a sense that the above eigenstates are orthogonal \cite{quasi1}. Such non-hermitian Hamiltonians are physically meaningful and are called quasi-hermitian Hamiltonians \cite{bender1}. Regardless of different physical motivations behind quasi-hermitian Hamiltonians, using the above quasi-hermitian Hamiltonian is important for us because we will be able to consider phase structure of $|K_d (\beta)\ra$ as a quantum phase transition in a model Hamiltonian. In particular, since $|K_d (\beta)\ra$ is the ground state of the above quasi-hermitian Hamiltonian, we can compute the ground state fidelity as a measure for characterizing a quantum phase transition.

  \section{Ground state fidelity and Mapping to d-state clock model}\label{sec3}
  As we mentioned in the previous section, by a quantum Hamiltonian with an exact ground state a simple way for characterizing quantum phase transition is to compute the ground state fidelity. The ground state fidelity is defined in the form of inner product of two consecutive ground states $|K_d (\beta) \ra$ and $|K_d (\beta + \delta \beta)\ra$ in the form of $F=\la K_d (\beta)|K_d (\beta +\delta \beta) \ra$. In order to compute this quantity, note that the normalization factor in the $|K_d (\beta) \ra$ should be in the form of $1/(\la K_d |\exp \{\beta \sum_i (Z_i +Z_i ^{-1})\}| K_d \ra )^{\frac{1}{2}}$. If we denote $\la K_d |\exp \{\beta \sum_i (Z_i +Z_i ^{-1})\}| K_d \ra$, which is a function of $\beta$, by $\mathcal{Z}(\beta)$, the ground state fidelity will find the following form:
\begin{equation}\label{ef}
F=\frac{\la K_d |\exp\{(\beta +\delta \beta /2)\sum_i (Z_i +Z_i ^{-1})\}|K_d \ra}{\sqrt{\mathcal{Z}(\beta)}\sqrt{\mathcal{Z}(\beta+\delta  \beta)}}
\end{equation}
Now, note that the numerator in the above equation is the same as $\mathcal{Z}(\beta +\delta \beta /2)$. Therefore, the  ground state fidelity will find the following simple form in terms of $\mathcal{Z}$:
\begin{equation}\label{ef2}
F=\frac{\mathcal{Z}(\beta +\delta \beta /2)}{\sqrt{\mathcal{Z}(\beta)}\sqrt{\mathcal{Z}(\beta+\delta  \beta)}}
\end{equation}
Now, we use a statistical mechanical mapping \cite{Nest2007} and prove that the function of $\mathcal{Z}$ is related to partition function of a classical d-state clock model.
\begin{figure}[t]
\centering
\includegraphics[width=5cm,height=5cm,angle=0]{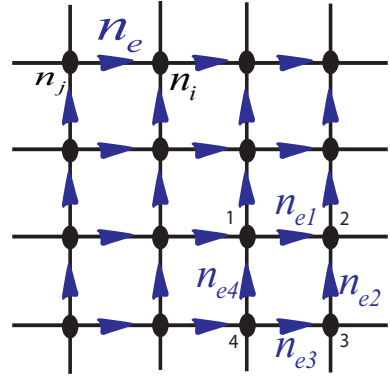}
\caption{D-state clock model defined on a square lattice with d-state variables living in vertices. new edge variables of $n_e$ are defined corresponding to each edge of the lattice. in the form of $n_e =n_i -n_j $. } \label{clock}
\end{figure}
To this end, consider a classical clock model on a square lattice where d-state variables of $\theta = 2\pi n/d ~;~n=0,1,..,d-1$ live in vertices and the classical Hamiltonian is in the following form:
\begin{equation}
H_{cl}=-\sum_{\la i,j\ra} \cos (\theta_i -\theta_j)
\end{equation}
where $\la i,j\ra$ refers to the interaction between the nearest neighbors. The partition function of such a system in terms of a finite temperature of $T$ is in the following form:
\begin{equation}
\mathcal{Z}_{clock}(T)=\sum_{\{\theta_i\}}\exp \{\sum_{\la i,j \ra} \cos (\theta_i -\theta_j)/T\}
\end{equation}
where we set the Boltzmann constant $k_B$ equalt to $1$ and the $\sum_{\{\theta_i\}}$ refers to summation on all configurations of d-state variables. Then if we use the fact that $\cos (x)=\frac{\exp\{ix\}+\exp\{-ix\}}{2}$ and using the equation $\omega=\exp \{i2\pi/d \}$, we will have:
\begin{equation}
\mathcal{Z}_{clock}(T)=\sum_{\{n_i\}}\exp \{\sum_{\la i,j \ra} \frac{\omega^{n_i -n_j }+\omega^{n_j -n_i}}{2T} \}
\end{equation}
where $n_i =0,1,...,d-1$ refer to different values of d-state variables of $\theta_i$. In the next step, we define new d-state variables corresponding to each edge of the square lattice in the form of $n_e =n_i -n_j $ which is called edge variables. We can replace these edge variables in the partition function relation. However, there is a point that new edge variables are not independent variables because of their definition in terms of $n_i$'s. For example consider a square plaquette of the lattice as shown in Fig.(\ref{clock}) where there are four vertex variables of $n_1 , n_2 , n_3 , n_4$. We consider a specific direction for each edge as shown in figure and accordingly each edge variable is equal to a vertex variable living in the end-point of the edge minus a vertex variable living in the first-point of that edge i. e. $n_{e1}=n_2 - n_1$, $n_{e2}=n_2 -n_3$, $n_{e3}=n_3 -n_4$ and $n_{e4}=n_1 -n_4$. By such a definition, it is concluded that there is a relation between edge variables corresponding to each plaquette in the form of $n_{e1}-n_{e2}-n_{e3}+n_{e4} =0$. Using the same notation that we had used for plaquette operators of Kitaev state in Sec.(\ref{sec1}), the above constraint can be  written in a compact form as $\sum_{e \in \partial p}\sigma_e n_e =0$. We apply these constraints by a set of delta functions in the partition function in the following form:
$$\mathcal{Z}_{clock}(T)=$$
\begin{equation}\label{ep}
\sum_{\{n_e \}}\exp \{\sum_{e} \frac{\omega^{n_e }+\omega^{-n_e} }{2T}\}\prod_{p}\delta(\prod_{e \in \partial p}\omega^{\sigma_e n_e} ,1 )
\end{equation}
On the other hand, since $\omega = \exp \{i2\pi /d \}$, it is simple to check that $1+\omega + \omega ^2 +...\omega^{d-1} =0$. Therefore, it is concluded that a delta function in the form of $\delta(\omega^{k},1)$ can be written in the form of $\frac{1}{d} \sum_{m=0}^{d-1} \omega^{km}$. In this respect, we rewrite each delta function in Eq.(\ref{ep}) in the following form:
\begin{equation}
\delta(\prod_{e \in \partial p}\omega^{\sigma_e n_e} ,1 )=\frac{1}{d}\sum_{m=0}^{d-1}(\prod_{e \in \partial p}\omega^{\sigma_e n_e})^m
\end{equation}
Now we are ready to introduce a quantum formalism for the partition function of (\ref{ep}). To this end, note that $\omega^{n_e}$ is in fact an eigenvalue of Pauli operator of $Z$ corresponding to eigenstate of $|n_e\ra$. Consequently, an arbitrary function in the form $\sum_{n_e} g(\omega^{n_e }) $ can be rewritten in the form of $ d\la + |g(Z)|+\ra$. Using the above equation we are able to write the partition function in the following quantum language:
$$\mathcal{Z}_{clock}(T)=d^{N}$$
\begin{equation}\label{ept}
    ~^{ N\otimes} \la +| \exp \{\sum_{e} \frac{Z_e +Z_e ^{-1} }{2T}\}\prod_{p}\frac{\sum_{m=0}^{d-1}(\prod_{e \in \partial p}Z_e ^{\sigma_e })^m}{d}|+\ra ^{\otimes N}
\end{equation}
Interestingly, the operator of $\prod_{e \in \partial p}Z^{\sigma_e}$ is the same as plaquette operator of $B_p$ in the $Z_d$ Kitaev model. Therefore, it is concluded that the state of $\prod_{p}\frac{\sum_{m=0}^{d-1}B_p ^m}{d} |+\ra ^{\otimes N}$ is the same as $Z_d$ Kitaev state up to a normalization factor where it is stabilized by all $B_p$ and $A_v$ stabilizer operators. On the other hand, since $\prod_{p}\frac{\sum_{m=0}^{d-1}B_p ^m}{\sqrt{d}}$ is a projective operator, it is concluded that it is equal to $(\prod_{p}\frac{\sum_{m=0}^{d-1}B_p ^m}{\sqrt{d}})^2$. Finally, by replacing in Eq.(\ref{ept}),  the partition function will be equal to an inner product in the following form:
\begin{equation}\label{ee}
\mathcal{Z}_{clock}(T)=d^{\frac{N}{2}}  \la K_d|  \exp \{\sum_{e} \frac{Z_e+Z_e ^{-1}}{2T}\}|  K_d \ra
\end{equation}
The above equation is exactly the same relation that we looked for.  If we go back to Eq.(\ref{ef}) and Eq.(\ref{ef2}) for the ground state fidelity, it would be concluded that the normalization factor of $\mathcal{Z}(\beta)$ in the ground state fidelity is the same as the partition function of clock model in Eq.(\ref{ee}) where $\frac{1}{2T}$ in the classical model has been mapped to $\beta$ in the ground state fidelity for the deformed Kitaev state. Furthermore, we should emphasize that the above mapping from the normalization factor of the deformed Kitaev state and partition function of the clock model is in fact a result of encoding Boltzmann weights of the clock model in amplitudes of the deformed Kitaev state. In other words, it is enough to write an expansion of the $Z_d$ Kitaev state in terms of string-nets. Then if we apply the LI operator $\exp \{\frac{\beta}{2} \sum_i (Z_i +Z_i ^{-1})\}$ to different terms of the above expansion, a square root of Boltzman weights of the clock model appear.

Finally, we rewrite the ground state fidelity in Eq.(\ref{ef2}) in the following form:
\begin{equation}\label{ew}
F_{K-state}(\beta, \delta \beta) =\frac{\mathcal{Z}_{clock} (\beta +\delta \beta /2)}{\sqrt{\mathcal{Z}_{clock}(\beta) \mathcal{Z}_{clock}(\beta +\delta \beta)}}
\end{equation}

It is in fact, a relation between the ground state fidelity in the quantum model and partition function of the clock model. On the other hand, since the parameter of $\delta \beta$ in the above equation is very small, it is useful to do a Taylor expansion for the state fidelity in terms of different powers of $\delta \beta$. By a simple calculation, we derive the following approximation for the ground state fidelity:
\begin{equation}
F_{K-state}(\beta, \delta \beta)\simeq 1- \frac{1}{8} (\frac{\partial ^2
\ln(\mathcal{Z}_{clock})}{\partial \beta^2})\delta \beta ^2
\end{equation}
where we have ignored higher powers of $\delta \beta$ . Furthermore, it is well-known that the second derivation of the partition function is related to the heat capacity for any classical statistical model. Therefore, we find that the ground state fidelity is related to the heat capacity of clock model by the following relation:
\begin{equation}\label{fid}
F_{K-state} (\beta ,\delta \beta)\simeq1-\frac{(C_v)_{clock}
}{8 \beta^2}\delta \beta ^2
\end{equation}
We remind that we considered the ground state fidelity to find a signature for the phase transition from $Z_d$ Kitaev phase to the trivial phase. Now we have found that this quantity is related to the partition function and the heat capacity of a classical clock model. It means that if the partition function or the heat capacity of the classical model show a singularity in a transition temperature of $T_c$, the ground state fidelity will also show a singularity in a $\beta_c =\frac{1}{ 2T_c}$ which will be indeed a topological phase transition point in our quantum model.

 \section{Identifying a KT-like quantum phase}\label{sec4}
In the previous section we have established a mapping from the deformed Kitaev model to the clock model. In particular we found that the ground state fidelity of the deformed Kitaev model is mapped to the partition function of the clock model Eq.(\ref{ew}). Then, by an expansion of the ground state fidelity in terms of different powers of $\delta \beta$, we showed that the ground state fidelity is related to the heat capacity of the clock model up to second order of $\delta \beta$ Eq.(\ref{fid}). By a mapping between the ground state fidelity for the deformed Kitaev model and the heat capacity of the clock model, now we are able to consider phase structure of our quantum system.

\begin{figure}[t]
\centering
\includegraphics[width=7cm,height=7cm,angle=0]{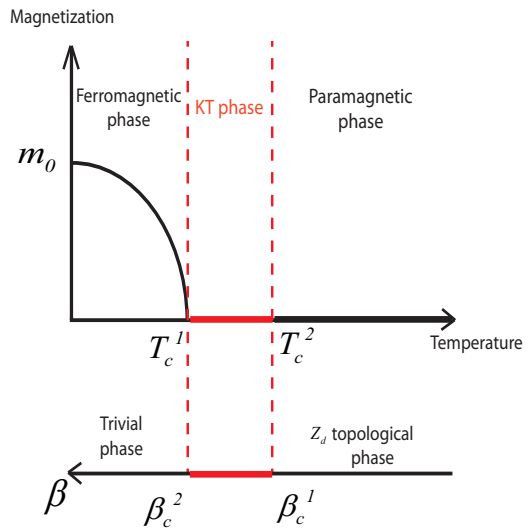}
\caption{A schematic of phase diagram of d-state clock model for $d \geq 5$ with two transition points of $T_c ^1$ and $T_c ^2$ where a KT phase emerges between ferromagnetic and paramagnetic phases. Although the magnetization is zero in the KT phase, the correlation function $C(r)$ shows a power law behavior where it decays in the form of $r^{-\eta}$. Since $\beta$ in the deformed Kitaev model is equal to $\frac{1}{2T}$, the trivial phase and the $Z_d$ topological phase correspond to ferromagnetic phase and paramagnetic phase, respectively.} \label{thouless}
\end{figure}

\begin{figure}[t]
\centering
\includegraphics[width=8cm,height=7cm,angle=0]{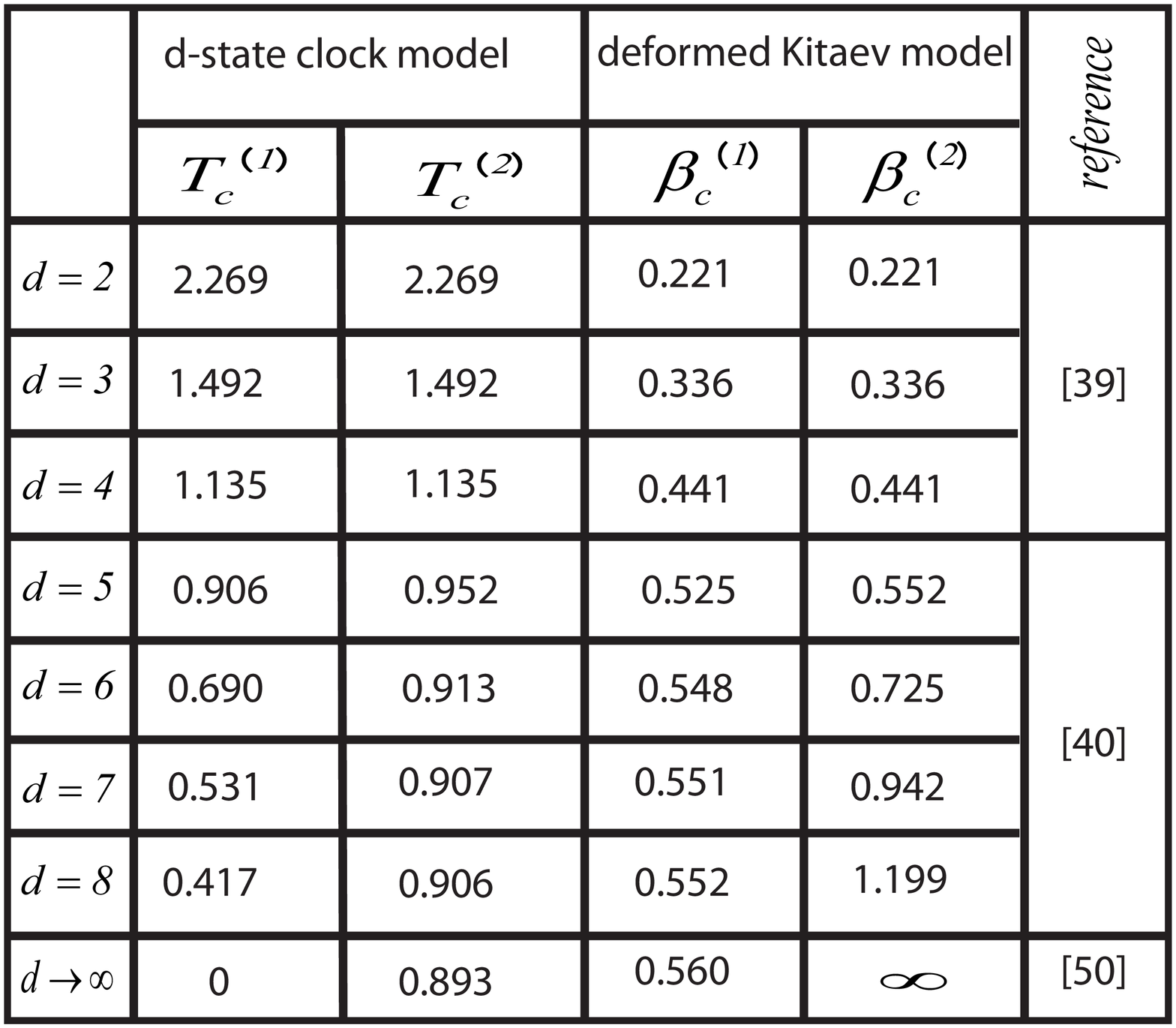}
\caption{According to denoted references, there are critical temperatures for d-state clock model for $d \leq 8$ and $d \rightarrow \infty$. Note that tempratures are dimensionless in the form of $\frac{k_B T}{J}$ and we have set $J$, $k_B$ equal to 1 where $J$ referes to the coupling constant of the clock model. The transition points of the quantum model are also determined by equations $\beta_c ^{1}=\frac{1}{2T_c ^{2}}$ and $\beta_c ^2 =\frac{1}{2T_c ^{1}}$ according to quantum-classical mapping.} \label{transition}
\end{figure}

 Fortunately, the clock model has well been studied in the literature in a sense that we have much information about its phase structure. Let us focus on the free energy and the heat capacity of this model which are related to the partition function in the form of $A=-k_B Tln(\mathcal{Z})$ and $C_v  = k_B \beta^2 \frac{\partial^2 ln \mathcal{Z}}{\partial \beta^2}$, respectively. It has been known that the clock model, for $d\geq 5$, shows two KT phase transitions \cite{clock00,clock1,clock0} where a KT phase emerges between ferromagnetic and paramagnetic phases, see Fig.(\ref{thouless}). Indicator of the above KT phase transitions is the existence of essential singularity in the free energy as a function of temprature at transition points denoted by $T_c ^{1}$ and $T_c ^{2}$, see Fig.(\ref{transition}) where values of transition tempratures have been given for a few values of $d$. Singular term of the free energy near transition points is in the form of $e^{-\frac{c}{t^{1/2}}}$ where $c$ is a positive constant and $t=\frac{T-T_c}{T_c}$ \cite{singular}. Such singularity is called essential because all derivatives of finite order of the free energy with respect to $T$ are continuous and there is only a divergence in derivative of infinite order. Furthermore, since the heat capacity is equal to second derivation of the free energy, it is also concluded that there is an essential singularity in the heat capacity as a function of $T$ at the KT transition point which is called a weak singularity \cite{singular, heat1,heat2}. 
 
 Now, using the mapping between the ground state fidelity and the partition function as well as the heat capacity, we conclude that there must be two singular points of $\beta_c ^{1} =\frac{1}{2T_c ^{2}}$ and $\beta_c ^{2} =\frac{1}{2T_c ^{1}}$ in the ground state of deformed Kitaev model where the ground state fidelity will show essential singularities as a function of $\beta$, see Fig.(\ref{thouless}). In this respect, we conclude that there are a trivial phase and a $Z_d$ topological phase in the quantum model corresponding to the ferromagnetic phase and the paramagnetic phase, respectively. However, for $d\geq 5$ there is not a simple phase transition from the $Z_d$ topological phase to the trivial phase but they are separated from each other by an intermediate quantum phase.

 Although singularities in the ground state fidelity can reveal different nature of the intermediate quantum phase, we need to characterize this phase in terms of some observable. To this end, note that the trivial phase, the intermediate phase and the $Z_d$ topological phase in the quantum model correspond to the ferromagnetic phase, the KT phase and the paramagnetic phase in the clock model, respectively. On the other hand, it has been known that three different phases in the d-state clock model can be characterized in terms of a correlation function in the form of $\la \cos (\theta_k -\theta _l ) \ra$ where $\theta_k $ and $\theta_l$ are two arbitrary d-state variables. This quantity is a function of $r$, distance between two the above variables in the lattice, and we denote it by $C(r)$. In the ferromagnetic phase $C(r)$ shows a long range order where it goes to a non-zero value in limit of $r \rightarrow \infty$ and it will be the same as ferromagnetic order parameter. In the paramagnetic phase, the system does not have a long range order and specifically $C(r)$ exponentially decays to zero. Interestingly, in the KT phase there is also no long-range order in a sense that the order parameter is equal to zero. However, the correlation function $C(r)$ decays to zero in an algebraical way where it shows a power law behavior in the form of $r^{-\eta}$ \cite{clock00}. In other words, the KT phase has a quasi long-range order.
 
 According to the above motivation, if we are able to find an observable in the quantum model corresponding to the correlation function in the clock model, such observable can be considered as signature for the intermediate phase in a sense that it reveals different behavior of the intermediate phase with trivial and $Z_d$ topological phases.

To this end, note that the correlation function of $\la \cos (\theta_k -\theta _l )\ra = \la \cos\frac{2\pi}{d}(n_k -n_l) \ra$ is equal to real part of $\exp\{i\frac{2\pi}{d}(n_k - n_l)\}$ which should be computed by the following relation:
$$\la \exp\{i\frac{2\pi}{d}(n_k - n_l)\} \ra =$$
\begin{equation}\label{deq}
\frac{\sum_{\{n_i\}}e^{i\frac{2\pi}{d}(n_k - n_l)}\exp\{ \sum_{\la i , j\ra}\frac{\cos \frac{2\pi}{d}(n_i - n_j)}{T}\}}{\mathcal{Z}}
\end{equation}
\begin{figure}[t]
\centering
\includegraphics[width=5cm,height=5cm,angle=0]{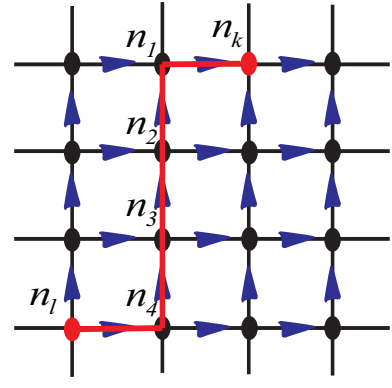}
\caption{A string of $S$, denoted by a red (light) color, connects two d-state variables of $n_k$ and $n_l$ where we consider a sequence of four d-state variables of $n_1$, $n_2$, $n_3$ and $n_4$ between them. By a mapping from classical vertex variables to quantum edge variables, the correlation function $\la \exp\{i\frac{2\pi}{d}(n_k - n_l)\} \ra$ is mapped to the expectation value of an operator of $ \prod_{e \in S} Z_e $ corresponding to the string of $S$} \label{correlation}
\end{figure}
Next, we rewrite the above equation in a quantum formalism by replacing vertex variables with edge variables. Everything is similar to the formalism that we used for partition function in Sec.(\ref{sec2}). However, here we have also another term of $e^{i\frac{2\pi}{d}(n_k - n_l)}$ which should be rewritten in terms of edge variables. To this end, as it is shown in Fig.(\ref{correlation}) for an example, we can consider a sequence, denoted by a string $S$, of d-state variables $n_1$ to $n_4$ between $n_k$ and $n_l$. Then we write $e^{i\frac{2\pi}{d}(n_k - n_l)}$ in the above equation in the following form:
\begin{equation}
 e^{i\frac{2\pi}{d}(n_k - n_1)} e^{i\frac{2\pi}{d}(n_1 - n_2)} e^{i\frac{2\pi}{d}(n_2 - n_3)} e^{i\frac{2\pi}{d}(n_3 - n_4)}e^{i\frac{2\pi}{d}(n_4 - n_l)}
\end{equation}
In this way, after change of variable, Eq.(\ref{deq}) is written in terms of edge variables $n_e$ in the following form:
$$
\la \exp\{i\frac{2\pi(n_k - n_l)}{d}\} \ra =
$$
\begin{equation}\label{deq2}
\frac{\sum_{\{n_e \}} \prod_{e \in S}e^{i\frac{2\pi n_e}{d} }e^{\sum_{e}\frac{\cos (\frac{2\pi n_e}{d} )}{T} }\prod_{p}\delta(\sum_{e\in p}\sigma^{e}n_e , 0) }{ \mathcal{Z}}
\end{equation}

Next, it is enough to replace variables of $\omega^{n_e}$ by Pauli operators of $Z$ by the same mechanism that we used for deriving Eq.(\ref{ee}). However, there is only a difference where we will have an operator in the form of $\prod_{e \in S} Z_e$ corresponding to  $\prod_{e \in S}e^{i(\frac{2\pi n_e }{d})}$ in the above equation. In this way, the correlation function will find the following form in the quantum language:
\begin{equation}\label{deq2}
\la e^{i(\theta_k -\theta_l)} \ra =\frac{\la K_d| (\prod_{e \in S} Z_e ) \exp \{\sum_{e} \frac{Z_e+Z_e ^{-1} }{2T}\}|  K_d \ra}{\la K_d|  \exp \{\sum_{e} \frac{Z_e+Z_e ^{-1} }{2T}\}|  K_d \ra}
\end{equation}
Interestingly, the above relation is in fact the same as expectation value of operator of $\prod_{e \in S} Z_e $, that we call it a string operator, in the deformed Kitaev state i. e. $\la K_d (\beta) |\prod_{e \in S} Z_e |K_d (\beta) \ra$. In this way, the expectation value of a string operator in the form of $\frac{\prod_{e \in S} Z_e + \prod_{e \in S} Z_e ^{-1}}{2}$, that we call it a string parameter, behaves similar to correlation function in the clock model. It means that if $r$ is the distance between two endpoints of the string $S$, the string parameter shows different behaviors as a function of $r$ for three different phases of the quantum model. In particular, while $r \rightarrow \infty$, it goes to a non-zero value in the trivial phase, it exponentially decays to zero for the $Z_d$ topological phase and finally it shows a power law behavior in the intermediate phase.

On the other hand, note that it has been shown that two quantum states which can be transformed to each other by small local invertible transformations are in the same topological class \cite{wen4}. Therefore, since the intermediate phase shows a singular transition to the trivial phase during the small local invertible transformation, it seems that the intermediate phase must be a topological (non-trivial) phase. However, note that the intermediate phase is also distinguished from the $Z_d$ topological phase by a power law behavior instead of exponential one. Finally, since the intermediate phase corresponds to KT phase in the clock model and there is also a power law behavior which is also observed in the correlation function in the KT phase, we call the intermediate phase a KT-like quantum phase.

We should emphasize that for $d \leq 4$, clock model shows a simple ferromagnetic-paramagnetic phase transition point. Specifically, for $d=2$ and $d=4$ there is a single critical point in the Ising universality class and for $d=3$ the clock model will be a three-state potts model with a single critical point. Therefore, for the corresponding quantum model the KT-like quantum phase disappears and we will have a simple topological phase transition from the $Z_d$ topological phase to the trivial phase.

Finally, we have also the phase transition points for the deformed Kitaev model according to well-known critical temperatures for clock model. In Fig.(\ref{transition}), according to a few recent papers in the literature, we have shown transition temperatures of clock model for $d\leq 8$ and the corresponding transition points for the deformed Kitaev model. Furthermore, clock model will be the same as X-Y model in $d \rightarrow \infty$ with a well-known phase transition point \cite{kt, xymodel}. In particular, in the X-Y model there is no ferromagnetic phase in finite temperature. On the other hand, the ferromagnetic phase in the classical model corresponds to the trivial phase of the quantum model. It means that for the deformed Kitaev model in $d \rightarrow \infty$, there are only a $Z_d$ topological phase and a KT-like phase where the trivial phase occurs in an infinite value of $\beta$.

It is also important to interpret the above transition points in the deformed Kitaev model as a measure of stability of topological phases. To this end, note that it is well-known that topological phases are stable against small LI transformations \cite{wen4}. Therefore, it is expected that $Z_d$ topological phase shows a stability against LI transformation that we considered in a sense that it must remain in the topological phase for small values of $\beta$. Consequently, we expect that by increasing $\beta$ the above stability breaks and a phase transition occurs. Therefore, we can interpret the $\beta_c  ^{1}$ as measure of stability of $Z_d$ topological phase. Furthermore, the $\beta_c ^{2}$ will also be interpreted as a measure of stability of KT-like quantum phase. By these interpretations and according to the table in Fig.(\ref{transition}), we conclude that the stability of both $Z_d$ topological phase and KT-like phase increases by increasing $d$. In particular, the KT-like phase shows an infinite stability for $d \rightarrow \infty$. It also means that the KT-like quantum phase is more stable than the $Z_d$ topological phase against the above LI transformation for large $d$.
\section{Discussion}
In mapping between classical spin models and quantum entangled systems, an interesting problem is to find quantum phase transitions corresponding to different well-known classical phase transitions. In this paper, we found an extended topological phase transition corresponding to Kosterlitz-Thouless phase transition in classical d-state clock model. In one hand, we mapped the ground state fidelity in a deformed Kitaev model to the heat capacity of the clock model and showed that there are three different phases in the deformed Kitaev model where an intermediate phase emerged between the $Z_d$ topological phase and the trivial phase. On the other hand, we mapped the correlation function in the clock model to a string parameter in the quantum model to characterize the intermediate phase in terms of an observable. We believe that the intermediate phase that we called a KT-like quantum phase might have important properties which have not been already seen in other quantum systems. In particular, while we know that the $Z_d$ topological phase has a long-range entanglement, it is important to consider entanglement in the KT-like quantum phase. Specifically, we note since the KT-like phase corresponds to a classical phase with a quasi long-range order, it might be a topological phase with a quasi-long range entanglement, a problem that should be considered in future works.
\section*{Acknowledgement}
We would like to thank A. Ramezanpour and A. Montakhab for their valuable comments.


\begin{thebibliography}{99}
\bibitem{sav}
S. Sachdev, \emph{Quantum Phase Transitions} (Cambridge University Press, Cambridge, UK, 2013)
\bibitem{entan}
L. Amico, R. Fazio, A. Osterloh, V. Vedral, Entanglement in many-body systems, Reviews of modern physics, 80(2), 517 (2008).
\bibitem{hamma}
A. Hamma, W. Zhang, S. Haas, D. A. Lidar, Entanglement, fidelity, and topological entropy in a quantum phase transition to topological order, Phys. Rev. B 77.15 (2008): 155111.
\bibitem{chen}
Y. Chen, and Sheng-Wen, Quantum correlations in topological
quantum phase transitions, Phys. Rev. A 81.3 (2010): 032120.
\bibitem{mont2010}
 A. Montakhab, A. Asadian, Multipartite entanglement and quantum phase transitions in the one-, two-, and three-dimensional transverse-field Ising
model, Phys. Rev. A 82, 062313 (2010).
\bibitem{fd}
S. Chen, L. Wang, Y. Hao, Y. Wang, Intrinsic relation between ground-state fidelity and the characterization of a quantum phase transition, Physical Review A, 77(3), 032111 (2008).
\bibitem{gu}
S. Gu, Fidelity approach to quantum phase transitions, Int. J. Mod. Phys. B 24.23 (2010): 4371-4458.
\bibitem{fidelity}
S. Yang, S. J. Gu, C. P. Sun, H. Q. Lin, Fidelity susceptibility and long-range correlation in the Kitaev honeycomb model, Phys. Rev. A 78.1 (2008): 012304.
\bibitem{Kitaev2003}
A. Y. Kitaev, Fault-tolerant quantum computation by anyons, Ann. Phys. (N.Y.) 303, 2 (2003).
\bibitem{rob1}
S. Trebst, P. Werner, M. Troyer, K. Shtengel, and C. Nayak, Breakdown of a topological phase: Quantum phase transition in a loop gas model with tension, Phys. Rev. Lett. 98, 070602 (2007)
\bibitem{rob2}
S. Dusuel, M. Kamfor, R. Orus, K. P. Schmidt, and J. Vidal, Robustness of a perturbed topological phase, Phys. Rev. Lett, 106, 107203, (2011).
\bibitem{zare16}
M. H. Zarei, Robustness of topological quantum codes: Ising perturbation, Phys. Rev. A 91, no. 2 (2015): 022319.
\bibitem{stringnet}
M. A. Levin, X. G. Wen, String-net condensation: A physical mechanism for topological phases, Physical Review B, 71(4), 045110 (2005).
\bibitem{zd}
G. Duclos-Cianci, D. Poulin, Kitaev's $Z_{d}$-code threshold estimates, Physical Review A, 87(6), 062338 (2013).
\bibitem{vidal1}
 L. Vanderstraeten, M. Mariën, J. Haegeman, N. Schuch, J. Vidal, F. Verstraete, Bridging perturbative expansions with tensor networks, Physical review letters, 119(7), 070401 (2017).
 \bibitem{vidal2}
A.  Schotte, J. Carrasco, B. Vanhecke, L. Vanderstraeten, J. Haegeman, F. Verstraete, J. Vidal, Tensor-network approach to phase transitions in string-net models, Physical Review B, 100(24), 245125 (2019).
\bibitem{del}
O. Viyuela, A. Rivas, M. A. Martin-Delgado, Generalized toric codes coupled to thermal baths, New Journal of Physics, 14(3), 033044 (2012).
\bibitem{karimi}
R. Mohseninia, S. S. Jahromi, L. Memarzadeh, V. Karimipour, Quantum phase transition in the Z 3 Kitaev-Potts model, Physical Review B, 91(24), 245110 (2015).
\bibitem{land}
L. D. Landau, Phys. Z. Sowjetunion. 11, 26 (1937),
Goldenfeld N., "Lectures on phase transitions and critical
phenomena", (Perseus Books Publishing, L.L.C., Massachusetts,
1992).
\bibitem{wwen}
X.-G Chen, G. Zheng-Cheng, and X.-G. Wen, Local unitary transformation, long-range quantum entanglement, wave function renormalization, and topological order, Phys. Rev. B 82.15 (2010): 155138.
\bibitem{wang}
H. Jiang, Z. Wang, and L. Balents, Identifying topological order
by entanglement entropy, Nat. Phys. 8.12 (2012): 902.
\bibitem{wen3}
M. Levin,  X. G. Wen, Detecting topological order in a ground
state wave function, Phys. Rev. Lett. 96(11), 110405 (2006).
\bibitem{kitpre}
A. Kitaev, and J. Preskill, Topological entanglement entropy, Phys. Rev. Let 96.11 (2006): 110404.
\bibitem{wen4}
B. Zeng, X. G. Wen, Gapped quantum liquids and topological order, stochastic local transformations and emergence of unitarity, Phys. Rev. B 91.12 (2015): 125121.
\bibitem{castel2005}
C. Castelnovo, C. Chamon, C. Mudry, P. Pujol, From quantum mechanics to classical statistical physics: Generalized Rokhsar-Kivelson Hamiltonian and the stichastic matrix form decomposition, Ann. Phys. 318(2), 316-344 (2005).
\bibitem{r1}
X. Chen, X. Yu, G. Y. Cho, B. K. Clark, E. Fradkin, Many-body localization transition in Rokhsar-Kivelson-type wave functions, Physical Review B, 92(21), 214204 (2015).
\bibitem{r3}
S. Papanikolaou, E. Luijten, E. Fradkin, Quantum criticality, lines of fixed points, and phase separation in doped two-dimensional quantum dimer models, Physical Review B, 76(13), 134514 (2007).
\bibitem{Somma2007}
R. D. Somma, C. D. Batista, G. Ortiz, Quantum approach to classical statistical mechanics, Phys. Rev. Lett. 99, 030603 (2007).
\bibitem{zarei18}
M. H. Zarei, A. Montakhab, Dual correspondence between classical spin models and quantum CSS states, Phys. Rev. A 98, 012337 (2018).
\bibitem{Dennis2002}
 E. Dennis, A. Kitaev, A. Landahl, and J. Preskill, Topological quantum memory, J. Math. Phys. 43, 4452 (2002).
\bibitem{Katzgraber2009}
H. G. Katzgraber, H. Bombin, M. A. Martin-Delgado, Error threshold for color codes and random three-body Ising models, Phys. Rev. Lett. 103, 090501 (2009).
\bibitem{zareim19}
M. H. Zarei, A. Montakhab, Phase transition in a noisy Kitaev toric code model, Phys. Rev. A 99, 052312 (2019)
\bibitem{zareiab}
M. H. Zarei, A. Ramezanpour, Noisy Toric code and random-bond Ising model: The error threshold in a dual picture, Phys. Rev. A 100, 062313 (2019).
\bibitem{fer}
D. Felice, C. Cafaro, S. Mancini, Information geometric methods for complexity, Chaos: An Interdisciplinary Journal of Nonlinear Science, 28(3), 032101 (2018).
\bibitem{fer1}
C. Cafaro, P. M. Alsing, Information geometry aspects of minimum entropy production paths from quantum mechanical evolutions, Physical Review E, 101(2), 022110 (2020).
\bibitem{castel}
D. F. Abasto, A. Hamma, P. Zanardi, Fidelity analysis of topological quantum phase transitions, Physical Review A, 78(1), 010301 (2008).
\bibitem{zarei19}
M. H. Zarei, A. Montakhab, classical criticality establishes quantum topological order, arxiv:1907.06216 (2019).
\bibitem{kt}
J. M. Kosterlitz, D. J. Thouless, Ordering, metastability and phase transitions in two-dimensional systems, Journal of Physics C: Solid State Physics. 6 (7): 1181–1203 (1972).
\bibitem{clock00}
C. M. Lapilli, P. Pfeifer, C. Wexler, Universality away from critical points in two-dimensional phase transitions, Physical review letters, 96(14), 140603 (2006).
\bibitem{clock1}
G. Ortiz, E. Cobanera, Z.  Nussinov, Dualities and the phase diagram of the p-clock model, Nuclear Physics B, 854(3), 780-814  (2012).
\bibitem{clock0}
J. Chen et al, Phase transition of the q-state clock model: Duality and tensor renormalization, Chinese Physics Letters, 34(5), 050503 (2017).
\bibitem{clock2}
Z. Q. Li, L. P. Yang, Z. Y. Xie, H. H. Tu, H. J. Liao, T. Xiang, Critical properties of the two-dimensional $ q $-state clock model, arXiv preprint arXiv:1912.11416.
\bibitem{singular}
H. Nishimori, G. Ortiz, Elements of phase transitions and critical phenomena, OUP Oxford (2010).
\bibitem{heat1}
D. S. Greywall, P. A. Busch, Heat capacity of fluid monolayers of He 4, Physical review letters, 67(25), 3535 (1991).
\bibitem{heat2}
E. Sagi, E. Eisenberg, Topological phase transition in a discrete quasicrystal, Physical Review E, 90(1), 012105 (2014).
\bibitem{bender1}
C. M. Bender, S. Boettcher, Real Spectra in Non-Hermitian Hamiltonians Having
PT Symmetry, Phys. Rev. Lett. 80, 5243-5246 (1998).
\bibitem{bender2}
C. M. Bender, Making sense of non-Hermitian Hamiltonians, Rep. Prog. Phys. 70, 947-
1018 (2007).
\bibitem{mostafa}
A. Mostafazadeh, Pseudo-Hermiticity versus PT symmetry: The necessary condition for
the reality of the spectrum of a non-Hermitian Hamiltonian, J. Math. Phys. 43, 205-214
(2002).
\bibitem{quasi1}
F. M. Fernandez, Non-Hermitian Hamiltonians and similarity transformations, International Journal of Theoretical Physics, 55(2), 843-850 (2016).
\bibitem{string}
M. H. Zarei, Ising order parameter and topological phase transitions: Toric code in a uniform magnetic field, Phys. Rev. B 100, 125159 (2019).
\bibitem{qudit1}
D. Gottesman, A. Kitaev, J. Preskill, Encoding a qubit in an oscillator, Physical Review A, 64(1), 012310 (2001).
\bibitem{qudit4}
M. D. Schulz, S. Dusuel, R. Orus, J. Vidal, K. P. Schmidt, Breakdown of a perturbed topological phase, New Journal of Physics, 14(2), 025005 (2012).
\bibitem{qudit2}
S. Pirandola, S. Mancini, S. L. Braunstein, D. Vitali, Minimal qudit code for a qubit in the phase-damping channel, Physical Review A, 77(3), 032309 (2008).
\bibitem{qudit3}
C. Cafaro, F. Maiolini, S. Mancini, Quantum stabilizer codes embedding qubits into qudits, Physical Review A, 86(2), 022308 (2012).
\bibitem{Nest2007}
M. Van den Nest, W. D\"{u}r, H. J. Briegel, Classical spin models and the quantum-stabilizer formalism, Phys. Rev. Lett. 98, 117207 (2007).
\bibitem{xymodel}
I. Dukovski, J. Machta, L. V. Chayes, Invaded cluster simulations of the XY model in two and three dimensions, Physical Review E, 65(2), 026702 (2002).
\end{thebibliography}
\end{document}